\documentstyle[12pt]{article}

\newcommand{\Qf}{\frac{{F'}^2}{F^2}}

\newcommand{\Df}{\frac{F''}{F}}

\newcommand{\q}{\frac{Q'}{Q}}
\newcommand{\Dq}{\frac{Q''}{Q}}

\newcommand{\f}{\frac{F'}{F}}
\newcommand{\ele}{\frac{L'}{L}}
\begin{document}
\title{Stationary and Axisymmetric Perfect Fluids with one
Conformal Killing Vector}
\author{Marc Mars$^\ast$ \\
School of Mathematical Sciences, \\
Queen Mary and Westfield College, \\
Mile End Rd, London, E1 4NS, U.K.\\
and \\ 
Jos\'e M. M. Senovilla\thanks{Also at Laboratori de F\'{\i}sica Matem\`atica, 
IEC, Barcelona.} \\
Departament de F\'{\i}sica Fonamental, Universitat de Barcelona, \\
Diagonal 647, 08028 Barcelona, Spain.}
\maketitle
\begin{abstract}
We study the stationary and axisymmetric non-convective differentially
rotating perfect-fluid solutions of Einstein's field equations
admitting one conformal symmetry. We analyse the two inequivalent Lie algebras
not exhaustively considered in \cite{S1} and show that 
the general solution for each Lie
algebra depends on one arbitrary function of one of the coordinates
while a set of three ordinary differential equations for four unknowns
remains to be solved. The conformal Killing vector of these solutions is
necessarily homothetic. We summarize in a table all the possible solutions
for all the allowed Lie algebras and also add a corrigendum to
an erroneous statement in \cite{S1} concerning the differentially
rotating character of one of the solutions presented.
\end{abstract}
\newpage

\section{Introduction}

In a recent paper \cite{S1} we considered the study of stationary
(non-static) and axisymmetric non-convective
differentially rotating perfect-fluid solutions
admitting one proper conformal Killing vector.
A purely geometric result \cite{S2} for stationary and
axisymmetric
spacetimes (with a well-defined axis of symmetry) 
admitting one conformal motion
implies that both the stationary and the axial Killing vectors
necessarily commute with the conformal Killing. Thus, the Lie algebra of the
three dimensional conformal group can only adopt four inequivalent forms, which
were labeled in \cite{S1} as Abelian Case and Cases I, II and III, and which
correspond to
\begin{eqnarray}
\mbox{\bf Abelian Case} \hspace{3cm} & &
\left [\vec{\xi},\vec{k} \right ] =\vec{0},  \nonumber \\
\mbox{\bf Case I} \hspace{3cm} & &
\left [ \vec{\xi},\vec{k} \right ] = b \vec{k}, \label{Lie}  \\
\mbox{\bf Case II} \hspace{3cm} & &
\left [\vec{\xi},\vec{k} \right ] = c \vec{\xi}, \nonumber \\
\mbox{\bf Case III} \hspace{3cm} & &
\left[ \vec{\xi},\vec{k} \right ] = v \vec{\eta}, \nonumber
\end{eqnarray}
where $\vec{\xi}$ is the timelike Killing vector, $\vec{\eta}$ is the axial
Killing vector, $\vec{k}$ is the conformal Killing and $b$, $c$ and $v$ are
arbitrary non-vanishing constants (these constants can be normalised to one,
but as they usually carry dimensions we prefer to retain them). In
\cite{S1}  the general solution of Einstein's field
equations for the Abelian and Case I was found. The general solution for the
Abelian Case depends on an arbitrary function of a single variable
and the perfect fluid satisfies the barotropic equation of state
$\rho= p +$const. The
solution is Petrov type D and the fluid velocity vector lies in the two
plane generated at each point by the two repeated principal directions
of the Weyl tensor. This solution was first found under completely different
hypotheses by one of us \cite{S3}. In Case I, the general solution is an
explicit Petrov type D metric with the fluid velocity vector lying outside the
two-plane generated by the repeated principal directions and the perfect fluid
satisfying the barotropic equation of state $\rho +3p=0$. Let us remark here
that \cite{S1} contained an erroneous statement concerning this solution which
is corrected at the end of this paper: the fluid velocity vector
for this solution was claimed to rotate differentially when the solution is,
actually, rigidly rotating. Then, as a consequence of the results in \cite{Sen},
this solution is contained in the general Wahlquist family \cite{Wa}
of Petrov type D rigidly rotating perfect fluids with equation of state
$\rho+3p=$const.

The Cases II and III were also considered, but the study of solutions was only
performed under the severe restriction imposed by an Ansatz of
separation of variables in one of the metric functions. For the first two
cases in (\ref{Lie}) this separation
of variables is a consequence of one of the Einstein field
equations but, for the remaining Cases II and III,
it had to be imposed as an additional
assumption in order to handle the field equations.
The solution under this
assumption was found and the conformal Killing vector turned out to be, in fact,
homothetic.

The aim of this paper is to complete the study performed in \cite{S1}
by analysing the general non-convective differentially
rotating perfect-fluid solution of Einstein's field equations for a stationary
and axisymmetric spacetime with one conformal symmetry such that the Lie
algebra of the conformal group belongs to one of the Cases II or III
above.
It turns out that there exists one perfect-fluid family of solutions
for both cases. The dependence 
of all metric coefficients of these solutions on one of the coordinates
is explicit. The dependence on the other non-trivial coordinate can be obtained
by solving a remaining compatible system of ordinary differential equations. 
The general solution of this system depends on an arbitrary function which
cannot be reabsorbed by any coordinate change. It also results that
the conformal Killing for these two families is necessarily homothetic which
proves that no non-convective stationary and axisymmetric perfect fluid
admitting one {\it proper} conformal motion exists apart from those in
\cite{S1},\cite{S3}. We remark that the families
we will find in this paper contain as particular cases the solutions for
Cases II and III considered in \cite{S1}.

The plan of the paper is as follows. In section 2 we recall the
line-element and the form of the conformal Killing for the Lie algebras II
and III. We rewrite the metric in a way which allows us to consider the two
cases together. Then we write the Einstein field equations for a perfect-fluid
energy-momentum tensor and concentrate our efforts in a subsystem constituted
by two partial differential equations for a single function $\Psi(x,y)$ (the
conformal factor of the metric). In section 3 we find the general solution of
this system of equations which turns out to be a set of four different families
that we label as Solutions 1, 2, 3 and 4. Section 4 is devoted to impose
the remaining Einstein equation (which is
quadratic in the second derivatives) for each of the four solutions. We 
prove that Solution 1 does not represent a perfect fluid.  
Regarding
Solutions 2 and 3, we show that they can be treated together and that
the quadratic equation gives rise to a system of ordinary differential
equations which can be shown to give no solution with a regular axis
of symmetry. In an Appendix we proof this statement completely for the Case
III and give the
main indications regarding the Case II (the proof for this case is much longer
but still not difficult to perform).
Let us remark here that the assumption of regular axis is not an
additional one in our work because the restriction to the four possible Lie
algebras (\ref{Lie}) is based essentially on this regularity of the axis
\cite{S1}-\cite{S2}.

Then we consider Solution 4, which
makes the quadratic equation compatible thus giving perfect-fluid solutions
of Einstein's equations which are analysed in Section 5, where
we rewrite the line-elements for
these solutions in the simplest forms and write down the remaining ordinary
differential equations including a discussion on the compatibility of the
system.
We also give the expressions for the density and pressure and for the
fluid velocity vectors. We finish the section by enclosing a
table which summarizes all the non-convective
stationary and axisymmetric perfect-fluid solutions of Einstein's equations
admitting one conformal Killing vector.
We end this paper by including
a corrigendum for an erroneous statement in \cite{S1}
concerning the solution in Case I (see above).

\section{Canonical forms of the metric and Einstein field equations}

As we have stated in the Introduction, we are interested in studying the
Einstein field equations for a non-convective, differentially
rotating perfect fluid in stationary and axisymmetric spacetimes admitting
one conformal motion such that the conformal algebra belongs to Cases II
or III defined above. The canonical form of the metric in these two 
cases is explicitly given in \cite{S1} where it is shown that there
exist coordinates in which the line-elements and the conformal
Killing vectors take the forms

\vspace{5mm}

\noindent {\bf Case II}
\begin{eqnarray}
ds^2= \frac{1}{\Psi^2(x,y)}\left[ -\tilde{F}(x)\left( \frac{}{} e^{-c y}
dt+ \tilde{P}(x) d\phi \right )^2+
\frac{\tilde{Q}^2(x)}{\tilde{F}(x)}d\phi^2+dx^2+dy^2 \right ],
\label{metrII}
\end{eqnarray}
\begin{eqnarray*}
\vec{k}= c t \frac{\partial}{\partial t} + \frac{\partial}{\partial y}.
\end{eqnarray*}

\noindent {\bf Case III}
\begin{eqnarray}
ds^2= \frac{1}{\Psi^2(x,y)}\left[ -F(x) dt^2
+ \frac{Q^2(x)}{F(x)} \left (\frac{}{}d\phi + \left(P(x)-v y\right ) dt \right 
)^2
+dx^2+dy^2 \right ], \label{metrIII}
\end{eqnarray}
\begin{eqnarray*}
\vec{k}= v t \frac{\partial}{\partial \phi} + \frac{\partial}{\partial y}.
\end{eqnarray*}
In both metrics the dependence of the function $\Psi(x,y)$
on the variable $y$ must be non-trivial because, otherwise,
the conformal Killing $\vec{k}$ becomes a true Killing field.

As we will see in what follows, it is very convenient to study the two Cases
II and III together as they have similar properties. It is obvious that the two
line-elements can be fused into the following expression
\begin{eqnarray}
ds^2= \frac{1}{\Psi^2(x,y)} \left [ - F(x) e^{-2cy}dt^2 + \frac{Q^2(x)}{F(x)}
\left ( d\phi + e^{-cy} \left (P(x)-vy \right ) dt \right ) ^2 + 
dx^2+dy^2 \right ], \label{metr}
\end{eqnarray}
where the expressions relating the functions $\tilde{F}(x)$, $\tilde{P}(x)$
and $\tilde{Q}(x)$ with $F(x)$, $P(x)$ and $Q(x)$ for Case II can be easily
obtained. Case III is recovered
by setting $c=0$ and Case II corresponds to $v=0$.
Thus, we can set the product $cv=0$ wherever it occurs in the calculation
of the Einstein tensor. 
The orthonormal tetrad that we will use in order to write
the components of the tensor quantities is given by
$$
\mbox{\boldmath$\theta^0$}=\frac{1}{\Psi}\sqrt{F}e^{-cy}\mbox{\boldmath $dt$},
\hspace{2mm}
\mbox{\boldmath$\theta^1$}=\frac{1}{\Psi} \frac{Q}{\sqrt{F}}\left (
\frac{}{} \mbox{\boldmath
$d\phi$} + e^{-cy} \left ( P - vy \right )  \mbox{\boldmath $dt$} \right ), 
\hspace{2mm} \mbox{\boldmath$\theta^2$}=\frac{1}{\Psi}\mbox{\boldmath$dx$},
\hspace{2mm}
\mbox{\boldmath$\theta^3$}= \frac{1}{\Psi}\mbox{\boldmath$dy$}.
$$
We are interested in 
perfect-fluid solutions of the Einstein field equations, so that the
energy-momentum tensor takes the following standard form
\begin{eqnarray*}
T_{\alpha\beta}= \left ( \rho + p \right ) u_\alpha u_\beta + p g_{\alpha\beta},
\end{eqnarray*}
where $\rho$ and $p$ are the energy density and the pressure of the
fluid respectively and $\vec{u}$ is the fluid velocity vector. The fluid motion
is taken as non-convective, that is to say, the velocity vector $\vec{u}$
lies on the two-plane generated at each point by the two Killing vectors
$\vec{\xi}$ and $\vec{\eta}$.
In the above orthonormal tetrad the fluid one-form 
$\mbox{\boldmath$u$}$ is given by
\begin{eqnarray*}
\mbox{\boldmath$u$}=u_0 \mbox{\boldmath$\theta^0$}+ u_1 \mbox{\boldmath$
\theta^1$}, \hspace{15mm} u_{0}^2-u_{1}^2=1,
\end{eqnarray*}
so that the Einstein field equations, in $c= 8 \pi G = 1$ units, read
\begin{eqnarray*}
S_{00}  = \left(\rho + p\right ) u_0^2 - p, \hspace{1cm}
S_{01}  = \left ( \rho + p \right ) u_0 u_1, \hspace{1cm}
S_{11}  = \left ( \rho + p \right ) u_1^2 +p, \\
S_{22} = S_{33}  = p, \hspace{1cm}
S_{02}=S_{03} = S_{12}=S_{13}=S_{23} = 0,
\end{eqnarray*}
where $S_{\alpha\beta}$ stands for the components of the Einstein tensor in the
$\{\mbox{\boldmath${\theta}^{\alpha}$}\}$ cobasis. Some of these
equations can be rewritten involving only the Einstein tensor,
while the energy density, pressure and velocity vector are
calculated once these equations are solved.  Due to the separation
of the metric into two orthogonal blocks, some of the Einstein tensor
components are identically vanishing in the tetrad we are considering.
Specifically, it can be shown by direct calculation that
\begin{eqnarray*}
S_{02}\equiv 0, \hspace{1cm} 
S_{03}\equiv 0, \hspace{1cm} S_{12}\equiv 0, \hspace{1cm}
S_{13}\equiv 0
\end{eqnarray*}
and consequently, the only non-trivially satisfied Einstein equations are
\begin{eqnarray}
\left ( S_{00}+S_{22} \right ) \left ( S_{11} - S_{22} \right ) -
S_{01}^2 = 0,\label{a} \\
S_{22} - S_{33} =  0, \label{b}\\
S_{23} = 0. \label{c}
\end{eqnarray}
We will now concentrate on the last two equations (\ref{b})-(\ref{c}),
which read
(after dropping some global non-vanishing factors)
\begin{equation}
\partial_{xy}\Psi = H(x) \Psi, \hspace{1cm}
\partial_{xx} \Psi - \partial_{yy} \Psi = G(x) \Psi .  \label{hg}
\end{equation}
where the expressions for $H(x)$ and $G(x)$ are
\begin{eqnarray}
H(x)= \frac{1}{4} \left ( c P P' \frac{Q^2}{F^2} - c \frac{F'}{F} + v P'
\frac{Q^2}{F^2} \right), \hspace{3cm} \label{hhh}\\
G(x) = \frac{1}{4} \frac{{F'}^2}{F^2} - \frac{1}{2} \frac{F'Q'}{FQ}
-\frac{1}{4}{P'}^2 \frac{Q^2}{F^2}+ \frac{1}{2} \frac{Q''}{Q}+
\frac{c^2}{4} \frac{P^2Q^2}{F^2} -\frac{c^2}{2} + \frac{v^2}{4} \frac{Q^2}{F^2},
\label{ggg}
\end{eqnarray}
the primes meaning derivatives with respect to $x$. These equations
were studied in \cite{S1} only under the simplifying assumption 
$\Psi = g(x) + h(y)$,
which obviously corresponds to the vanishing of the function $H$. Given that
this case was completed in the mentioned paper, we
can assume from now on that $H$ does not vanish identically. In the next
section we are going to find the general solution
of the system of partial differential equations (\ref{hg})
for the function $\Psi(x,y)$ when $H\neq 0$.

\section{Solution of the system of equations for $\Psi$.}

In order to solve the system (\ref{hg}), let us find all the third-order
partial derivatives of $\Psi$ with respect to the variables $x$ and $y$.
Deriving the first of the equations with respect to $x$ and with respect to $y$
we immediately get, respectively
\begin{eqnarray}
\partial_{xxy} \Psi = H' \Psi + H \partial_x \Psi, \hspace{1cm}
\partial_{xyy} \Psi = H \partial_y \Psi,  \label{four1}
\end{eqnarray}
and considering the derivatives of the second equation we find, after making
use of (\ref{four1}) 
\begin{eqnarray}
\partial_{xxx} \Psi = H \partial_y \Psi + G \partial_x \Psi + G' \Psi , 
\hspace{1cm}
\partial_{yyy} \Psi = H' \Psi + H \partial_x  \Psi - G \partial_y \Psi .
\label{four2}
\end{eqnarray}
We can now impose integrability conditions on this set of partial differential
equations. For instance, the difference of the partial derivative of the first
equation in (\ref{four1}) with respect to $x$ with the partial derivative of
the first equation in (\ref{four2}) with respect to $y$ gives the following
first order equation
\begin{eqnarray}
H'' \Psi + 2 H' \partial_x \Psi = G' \partial_y \Psi, \label{pri}
\end{eqnarray}
where the original system (\ref{hg}) has been taken into account.
It is easy to see that the integrability condition for the other two equations
is given again by (\ref{pri}).

If the function $G$ is a constant, then (\ref{pri}) implies that either $H$ is
also a constant or $\Psi$ takes the form of a product of a function
of $x$ and a function of $y$.
From (\ref{pri}) it also follows that when $H$ is a constant then
$G(x)$ must also be constant (because the other possibility
$\partial_y \Psi = 0$ needs not be considered as discussed above).
Thus, let us first study the case when both $H$ and $G$ are constants.
Since we can assume $H\neq 0$, the original system (\ref{hg}) implies the
following homogeneous equation 
\begin{eqnarray*}
\partial_{xx} \Psi - \partial_{yy} \Psi - \frac{G}{H} \partial_{xy} \Psi = 0
\end{eqnarray*}
which can be immediately solved (given that $G$ and $H$ are constants) to give
\begin{eqnarray*}
\Psi= f_1 \left( y -a^2 H x \right) + f_2 \left( y + \frac{1}{a^2 H} x \right)
\end{eqnarray*}
where $f_1$ and $f_2$ are arbitrary functions of their arguments and the
positive constant $a$ is defined by
\begin{equation}
a^2 \equiv \frac{-G + \sqrt{G^2 + 4 H^2}}{2H^2} \hspace{5mm} \Longrightarrow
\hspace{5mm} \frac{1}{a^2}=\frac{1}{2}\left(G+\sqrt{G^2 + 4 H^2}\right).
\label{GG} 
\end{equation}
With this form for the function $\Psi$, the second equation in (\ref{hg}) can
be trivially separated into the following two equations
\begin{eqnarray}
\ddot{f}_1 = -\frac{1}{a^2} \left(f_1 + \frac{\alpha}{2 H}\right), \hspace{1cm}
\ddot{f}_2 = H^2 a^2 \left ( f_2 - \frac{\alpha}{2H} \right ). \label{sysy}
\end{eqnarray}
where $\alpha$ is a separation constant and dots indicate derivatives with
respect to their respective arguments. The redefinitions
$f_1 +\frac{\alpha}{2 H} \longrightarrow f_1$ and
$f_2 - \frac{\alpha}{2H} \longrightarrow f_2$, which
leave invariant its sum $\Psi =f_1+f_2$, allow us
to set the constant $\alpha = 0$. Thus, the solution of the system (\ref{sysy})
is clearly
\begin{eqnarray*}
\mbox{{\bf Solution 1}:} \hspace{1cm}
\Psi = A e^{aHy + \frac{x}{a}} + B e^{-\left ( aHy + \frac{x}{a} \right )}
+ C \sin \left ( \frac{y}{a} - aHx + x_0 \right )
\end{eqnarray*}
where $A$, $B$, $C$ are arbitrary constants and $x_0$ is a superfluous
constant that can be reabsorbed into the coordinates. Let us remember that
$G$ is obtained in terms of $H$ and $a$ from (\ref{GG}).

We shall momentarily discard the case when $G$ is a constant and $\Psi$ is a
product of a function of $x$ and a function of $y$ because it is contained in
a more general situation we will meet later. Let us then
consider the general case when $G'\neq 0$ (and then $H'\neq 0$).
Deriving equation (\ref{pri}) with respect to $x$ and using the first equation
in (\ref{hg}) we get 
\begin{eqnarray}
\frac{2H'}{G'} \partial_{xx} \Psi + \left [ \left ( \frac{2H'}{G'} \right )'
+ \frac{H''}{G'} \right ] \partial_x \Psi + \left [ \left ( \frac{H''}{G'} 
\right )' - H \right ] \Psi =0. \label{ord}
\end{eqnarray}
The nice property of this partial differential equation is that all the
coefficients and all the derivatives involve only the variable $x$.
Thus, its general solution is
\begin{eqnarray}
\Psi =  A_1(y) v_1(x) + A_2(y) v_2(x), \label{form}
\end{eqnarray}
where $A_1$ and $A_2$ are arbitrary integration coefficients which can depend on
$y$, and $v_1$ and $v_2$ are two linearly independent solutions (so that in
particular they are non-vanishing) of the ordinary second
order differential equation given by the same expression (\ref{ord}) with
ordinary derivatives substituted for the partial derivatives.
Our aim now is to find all the solutions of the original system (\ref{hg}) with
the form for the function $\Psi$ given in (\ref{form}). It is obvious that we
will also find all the solutions which are products of a function of $y$ and
a function of $x$ (say by imposing $A_2(y)=0$ in (\ref{form})), and thus the
case $G'=0$ with $H' \neq 0$ we had left aside is also
contained in this study.

The first equation in the system (\ref{hg}) for a function $\Psi$ of the type
(\ref{form}) takes the form
\begin{eqnarray}
\dot{A_1} v_1' + \dot{A_2} v_2' - A_1 H v_1 - A_2 H v_2 = 0, \label{fourA}
\end{eqnarray}
where dots mean derivative with respect to $y$ from now on.
The lefthand side of this equation is a linear combination of the four
functions of $y$, $\dot{A_1}$, $\dot{A_2}$, $A_1$ and $A_2$, with coefficients
not depending on $y$. Thus, we can study this equation by considering all the
possibilities of linear independence of these four functions.
First of all, it is obvious that the case in which all four functions of
$y$ are linearly independent need not be considered as it implies the
vanishing of each coefficient in (\ref{fourA}) which is clearly impossible.
Thus, we can move to the case in which $A_1$ and $A_2$ are 
linearly independent and $\mbox{rank} \{ \dot{A_1}, \dot{A_2}, A_1, A_2 \} = 3$,
so that without loss of generality we can consider that $\dot{A_2}$ is
linearly independent of $A_1$ and $A_2$ and write
$\dot{A_1} = \beta_1 \dot{A_2} + \beta_2 A_1 + \beta_3 A_2$,
where $\beta_1$, $\beta_2$, $\beta_3$ are constants.
The equation (\ref{fourA}) becomes
\begin{eqnarray*}
A_1 \left( \beta_2 v_1' - H v_1 \right) + A_2 \left ( \beta_3 v_1'
 - H v_2 \right ) + \dot{A_2} \left ( \beta_1 v_1' + v_2' \right ) =0,
\end{eqnarray*}
so that each term between brackets must vanish. From the coefficient multiplying
$A_1$ and $H\neq 0$ we have that $\beta_2\neq 0$ and thus
$v_1' = H v_1/\beta_2$. Substituting this value of $v_1'$ into the second term
in brackets and dropping a global factor $H$ we obviously get
$v_2 = \beta_3 v_1/\beta_2$ so that the two functions $v_1$ and $v_2$ are
linearly dependent which is impossible. Thus, this possibility
gives no solution for the function $\Psi$ either.

Let us now consider the next possibility given by $A_1$ and $A_2$ 
linearly independent and $\mbox{rank} \{ \dot{A_1}, \dot{A_2}, A_1, A_2 \} = 2$.
Thus, each function $\dot{A_1}$ and $\dot{A_2}$ is a linear combination with
constant coefficients of $A_1$ and $A_2$:
\begin{eqnarray*}
\dot{A_1} = \beta_1^1 A_1 + \beta_1^2 A_2,\hspace{1cm}
\dot{A_2} = \beta_2^1 A_1 + \beta_2^2 A_2.
\end{eqnarray*}
We can redefine the two functions $A_1$ and $A_2$ using
arbitrary non-singular linear combinations of themselves
because this does not change the form (\ref{form}) of $\Psi$ (obviously,
we must redefine the two functions $v_1$ and $v_2$ appropriately). Under such a
change, the coefficients $\beta_a^b$ ($a,b$ = 1,2) transform like an
endomorphism that can be rewritten in its canonical form. Consequently, the only
two inequivalent possibilities that can arise are
\begin{equation}
\left . 
\begin{array}{l}
\displaystyle{\dot{A_1} = \beta_1 A_1 }\\ 
\displaystyle{\dot{A_2} = \beta_2 A_2.} 
\end{array} \right \}, \hspace{2cm} \left . \begin{array}{l}
                           \displaystyle{ \dot{A_1} = \beta A_1} \\ 
                           \displaystyle{ \dot{A_2} = A_1 + \beta A_2 }
                                            \end{array} \right \} \hspace{2mm} .
                     \label{twoset}
\end{equation}
In the second case, equation (\ref{fourA}) takes the form
\begin{eqnarray*}
A_1 \left ( \beta v_1' - H v_1 + v_2' \right ) + A_2 \left (\beta v_2' - H v_2
\right ) = 0
\end{eqnarray*}
which implies that each term enclosed in round brackets must vanish.
Again $\beta$ must be non-vanishing so that we can set
$\beta = \frac{1}{\alpha}$. The solutions for $v_1$ and $v_2$ are
trivially
\begin{eqnarray*}
v_1 = r(x)  e^{\alpha r(x)},\hspace{1cm}
v_2 = - \frac{1}{\alpha^2} e^{\alpha r(x)}, \hspace{7mm} \mbox{with}
\hspace{7mm} r' = H,
\end{eqnarray*}
while the solutions for $A_1$ and $A_2$ are
\begin{eqnarray*}
A_1 = A e^{\frac{y}{\alpha}}, \hspace{1cm} A_2= A \left (y - y_0 \right )
e^{\frac{y}{\alpha}},
\end{eqnarray*}
where $A\neq 0$ and $y_0$ are constants. Obviously, we can reabsorbe $y_0$
into the variable $y$. Finally, by imposing the second equation in (\ref{hg})
and using the above results we obtain the following two equations
\begin{eqnarray}
r'' + 2 \alpha r'^2 + \frac{2}{\alpha^3} = 0,\hspace{1cm}
G  + \alpha^2 {r'}^2 + \frac{3}{\alpha^2}= 0. \label{relatC}
\end{eqnarray}
The first equation gives $r(x)$ explicitly (and thereby the
function $H(x)$ after derivation), and the second gives $G(x)$. Thus,
both of them are differential relations for the unknown metric coefficients
$F(x)$, $Q(x)$ and $P(x)$. Summing up, we have 
\begin{eqnarray*}
\mbox{{\bf Solution 2}:} \hspace{1cm}
\Psi = A e^{\alpha r(x) + \frac{y}{\alpha}}\left (r(x) - 
\frac{y}{\alpha^2} \right ) \hspace{10mm} \mbox{with} \hspace{4mm} H=r'
\end{eqnarray*}
and the two relations (\ref{relatC}) must hold.

\vspace{3mm}

We must now consider the possibility given by the first set of differential
relations in (\ref{twoset}). Using these equations, the relation (\ref{fourA})
takes the form
\begin{eqnarray*}
A_1 \left ( \beta_1 v_1' - H v_1 \right ) + A_2 \left ( \beta_2 v_2' - H v_2
\right ) = 0,
\end{eqnarray*}
so that again
each term in brackets must vanish. As neither $v_1$ nor $v_2$ can
be zero, we have that the constants $\beta_1$ and $\beta_2$ are non-vanishing
so that we can rename them respectively as $\frac{1}{\alpha_1}$ and
$\frac{1}{\alpha_2}$. The equations for $\dot{A_1}$ $\dot{A_2}$, $v_1'$ and
$v_2'$ allow us to write the final form of $\Psi$ immediately as
\begin{eqnarray*}
\mbox{{\bf Solution 3}:} \hspace{1cm}
\Psi = A e^{\alpha_1 r(x) + \frac{y}{\alpha_1}} + B e^{\alpha_2 r(x) + 
\frac{y}{\alpha_2}}, \hspace{1cm} H=r'
\end{eqnarray*}
(with $\alpha_1 \neq \alpha_2$ in order to have $A_1$ linearly independent of
$A_2$). The second equation of (\ref{hg}) is equivalent now to the two
equations:
\begin{eqnarray}
r'' + \left ( \alpha_1 + \alpha_2 \right ) {r'}^2 + \frac{\alpha_1+\alpha_2}{
\alpha_1^2 \alpha_2^2 } = 0, \hspace{6mm}
G + \alpha_1 \alpha_2 {r'}^2 + \frac{1}{\alpha_1^2} +
\frac{1}{\alpha_2^2} + \frac{1}{\alpha_1 \alpha_2} = 0. \label{Gr'1}
\end{eqnarray}
It is convenient to note here
that the constants $\alpha_1$, $\alpha_2$, $A$ and $B$ can be complex
still giving a real function
$\Psi$. This can be accomplished only when  
$\alpha_1 = \alpha + i \beta$, $\alpha_2= \overline{\alpha}_1$,
$A = \overline{B}$ where the bar means complex conjugation. In this case the
function $\Psi$ becomes 
\begin{eqnarray*}
\Psi = A e^{\alpha r(x) + \frac{\alpha}{\alpha^2+\beta^2}y} \sin \left ( \beta
r(x) - \frac{\beta}{\alpha^2+ \beta^2} y + x_0 \right )
\end{eqnarray*}
where $x_0$ is a constant that can be made zero by redefining the variable $x$.

\vspace{3mm}

Let us finally consider the case when the two functions $A_1(y)$ and $A_2(y)$
are linearly dependent. Due to the symmetry in the indexes $1$ and $2$ we can
choose $A_2$ proportional to $A_1$, $A_2 (y) = \lambda A_1(y)$
where $\lambda$ is an arbitrary constant. The form of the function $\Psi$ is now
\begin{eqnarray*}
\Psi(x,y) = A_1(y) \left[v_1(x) + \lambda v_2(x) \right] \equiv A_1(y) X(x)
\end{eqnarray*}
so that it is the product of a function of only $x$ and a function of only $y$.
The first equation in (\ref{hg}) is now
\begin{eqnarray*}
X' \, \dot{A_1} = H(x) X A_1 \hspace{1cm} \Longrightarrow \hspace{1cm}
\dot{A_1} = \frac{1}{\alpha} A_1, \hspace{1cm} X' = \alpha H X 
\end{eqnarray*}
for some non-vanishing separation constant $\alpha$.
The second equation of the system (\ref{hg}) becomes (after dropping a
global factor $\Psi$)
\begin{eqnarray}
G(x) - \alpha r'' -  \alpha^2 {r'}^2 +  \frac{1}{\alpha^2} = 0. \label{apCG}
\end{eqnarray}
Thus, the final possibility for the system (\ref{hg}) is given by
\begin{eqnarray*}
\mbox{{\bf Solution 4}:} \hspace{1cm}
\Psi = A e^{\alpha r(x) + \frac{y}{\alpha}},  \hspace{3cm} r'=H
\end{eqnarray*}
together with the differential equation (\ref{apCG}).

\section{Analysis of the remaining Einstein equation}

Having found all the possible solutions of the system (\ref{hg}),
we must now impose the quadratic equation (\ref{a}) on each one of these
possibilities in order to find which of them give rise to a
perfect-fluid solution of Einstein's equations. 
In order to deal with the mentioned quadratic equation
it is worth considering it as the vanishing of the norm of a three-dimensional
vector with components
\begin{eqnarray}
\vec{S} \equiv \left ( S_{00}+S_{33}, S_{11}-S_{22}, S_{01} \right ) \label{vec}
\end{eqnarray}
in a three-dimensional vector space endowed with a scalar product given by
\begin{eqnarray*}
\left ( 
          \begin{array}{ccc}
	  0 & \frac{1}{2} & 0 \\
	  \frac{1}{2} & 0 & 0 \\
	  0 & 0 & -1 
	  \end{array}
\right ).
\end{eqnarray*}
This quadratic form has signature $\left ( 1, -1, -1 \right )$, and therefore,
equation (\ref{a}) states that the vector (\ref{vec}) is null in this
indefinite scalar product. Now, evaluating the components of the Einstein
tensor of the metric (\ref{metr}) we find 
\begin{displaymath}
\begin{array}{c}
S_{00}+S_{33}  = \Psi \left[ \frac{}{} R_1(x) \Psi + R_2(x) \Psi_{,x} + 2 
\Psi_{,yy} + 2 c \Psi_{,y} \right], \\
\\
S_{11}-S_{33} = \Psi \left[ \frac{}{} M_1(x) \Psi + M_2(x) \Psi_{,x}  - 2 
\Psi_{,yy} \right], \\
\\
 S_{01} = \Psi \left[ \frac{}{} L_1(x) \Psi + L_2(x) 
 \Psi_{,x} + L_3(x)  \Psi_{,y} \right]
\end{array}
\end{displaymath}
where the expressions for $R_1$, $R_2$, $M_1$, $M_2$, $L_1$, $L_2$
and $L_3$ are
\begin{eqnarray*}
R_1= \frac{1}{2} \left ( \frac{F''}{F} - \frac{{F'}^2}{F^2} + \frac{F'Q'}{FQ}
-{P'}^2 \frac{Q^2}{F^2} \right), \hspace{1cm} R_2= -\frac{F'}{F}, \hspace{17mm}\\
M_1 = R_1 - \frac{Q''}{Q} - c^2 \frac{P^2Q^2}{F^2} + c^2 - v^2 \frac{Q^2}{F^2},
\hspace{1cm} M_2 = 2\frac{Q'}{Q} - \frac{F'}{F}, \hspace{13mm}\\
L_1 = \frac{Q}{F} \left ( P' \frac{F'}{F} - \frac{1}{2} P'' -\frac{3}{2} P' 
\frac{Q'}{Q} \right) , \hspace{1cm} L_2= P' \frac{Q}{F}, \hspace{1cm} L_3= - 
\frac{Q}{F} \left (cP + v \right ).
\end{eqnarray*}
Let us now study the consequences of equation (\ref{a}) when the function
$\Psi$ takes each one of its possible forms. Let us begin by what we have
named Solution 1.

\paragraph{Solution 1}
As convenient notation let us define the following three functions of
the two variables $x$ and $y$
\begin{eqnarray*}
k_1(x,y) \equiv e^{aHy + \frac{x}{a}}, \hspace{6mm} k_2(x,y) \equiv \sin \left (
 \frac{y}{a}
- aHx \right ), \hspace{6mm} k_3(x,y) \equiv \cos \left ( \frac{y}{a}- aHx \right ),
\end{eqnarray*}
so that $\Psi$ is written as $\Psi = A k_1 + B/k_1 + C k_2$.
If in this expression $C$ vanished, the solution 
would in fact belong to Solution 3. Similarly, if both $A$
and $B$ vanished, the solution would be also contained in Solution 3.
In consequence, we must have $C\neq 0$ and at least one of the $A$ and $B$
non-zero. By using the change $a \rightarrow -a$ we can obviously assume
$A$ and $C$ non-vanishing in what follows without loss of generality .

By taking the derivatives of the function $\Psi$, the vector (\ref{vec}) becomes
\begin{equation}
\frac{\vec{S}}{\Psi}= A k_1 \vec{X_1} + \frac{1}{k_1} \vec{X_2}
+ C k_2 \vec{X_3} + C k_3 \vec{X_4} , \label{vec2}
\end{equation}
where the four vectors $\vec{X_j} \hspace{2mm} (i,j=1,2,3,4)$ depend only on $x$
and are given by
\begin{eqnarray*}
\vec{X_1} = \left (R_1 + \frac{R_2}{a} + 2 a^2 H^2 + 2 c a H,
M_1 + \frac{M_2}{a} - 2 a^2 H^2, L_1 + \frac{L_2}{a} + aH L_3 \right ), \\
\vec{X_2} = B \left ( R_1 - \frac{R_2}{a} + 2 a^2 H^2 - 2caH, M_1 -
\frac{M_2}{a} - 2 a^2 H^2, L_1 - \frac{L_2}{a} - aH L_3 \right ), \\
\vec{X_3} = \left (R_1 - \frac{2}{a^2}, M_1 + \frac{2}{a^2}, L_1 \right ), 
\hspace{3cm}\\
\vec{X_4} = \left ( \frac{2c}{a} - aH R_2, - aH M_2 , -aHL_2 + \frac{L_3}{a}
\right ), \hspace{25mm}
\end{eqnarray*}
and we have included $B$ in $\vec{X_2}$ because $B$ is the only constant
that can vanish. Taking into account that the {\it only} vanishing linear
combination of products of two functions in the set $\{
k_1, 1/k_1, k_2, k_3\}$ is $k_1 /k_1 - k_2^2 - k_3^2 = 0$, the vanishing of the
norm of the vector (\ref{vec2}) implies
\begin{eqnarray*}
& \displaystyle{\vec{X}_1 \cdot \vec{X}_1 = \vec{X}_1 \cdot \vec{X}_3 =
\vec{X}_1 \cdot \vec{X}_4 = \vec{X}_2 \cdot \vec{X}_2 =
\vec{X}_2 \cdot \vec{X}_3 = \vec{X}_2 \cdot \vec{X}_4 =
\vec{X}_3 \cdot \vec{X}_4 = 0,}& \\
& \displaystyle{ \vec{X}_3 \cdot \vec{X}_3 - \vec{X}_4 \cdot \vec{X}_4 = 0},
\hspace{15mm}
\displaystyle{ 2A\vec{X}_1 \cdot \vec{X}_2 + C^2 \vec{X}_4 \cdot \vec{X}_4 =0}.& 
\end{eqnarray*}
If $\vec{X}_1 \cdot \vec{X}_2 \neq 0$ then the vector $\vec{X}_4$ would be
non-null and $\left\{\vec{X}_1,\vec{X}_2,\vec{X}_4 \right\}$ would be a basis of
the vector space. From the first set of equations above it would follow
that $\vec{X}_3$ should be the zero vector. Then, the second equation would be
incompatible. Thus, we must necessarily have $\vec{X}_1 \cdot \vec{X}_2 =0$ and
every scalar product of each vector $\vec{X_i}$ with each $\vec{X_j}$ 
must vanish. Therefore, all these four vectors are proportional to some null
vector and, as an obvious consequence, we have that the linear combination
\begin{eqnarray*}
\vec{X_1} - \vec{X_3} + \frac{1}{a^2H^2} \vec{X_4} = 
\left[ a^2H^2 + \frac{1}{a^2} \right] \left ( 2+ 2\frac{c}{aH} , 
- 2 , \frac{L_3}{aH} \right )  
\end{eqnarray*}
has to be also a null vector, that is to say
\begin{eqnarray*}
4 + \frac{4c}{aH} + \frac{ {L_3}^2 }{a^2 H^2 } = 0,
\end{eqnarray*}
from which we learn that $c$ cannot vanish. Consequently, Case III is
impossible and we can set $v=0$, so that from the definition of $L_3$ we have
$L_3= - c \frac{QP}{F}$. The equation above shows that $L_3$
must be constant so that every vector $X_i$ is proportional to the constant
null vector $\left ( 2+ 2c/aH , - 2 , L_3/aH \right )$.
In particular, from the proportionality of $\vec{X_4}$ with
this vector and the fact that $H$ is actually a constant in Solution 1,
it immediately follows that $\frac{F'}{F}$ and $\frac{Q'}{Q}$ are also
constants. Now, it is very easy to find the most general solution of Einstein's
equations in this case, given by the following functions 
\begin{eqnarray}
\frac{Q'}{Q} = \frac{2}{a}, \hspace{3mm} P= \epsilon \frac{F}{Q} ,
\hspace{3mm} \frac{F'}{F} = \gamma
\hspace{3mm}  c= - 2aH, \hspace{3mm} \Psi =  A e^{aHy + \frac{x}{a}} + C 
\sin \left ( \frac{y}{a} - aHx \right ). \label{solp12}
\end{eqnarray}
where $\epsilon$ is a sign and $\gamma$ an arbitrary constant. The Einstein
tensor for the metric (\ref{metr}) with these explicit forms for the
functions $F$, $Q$, $P$ and $\Psi$ takes the following form
\begin{eqnarray*}
S_{00}+ S_{22} = S_{11}- S_{22} = - \epsilon S_{01}, \hspace{1cm}
S_{22} = S_{33} = 3 C^2 \left ( a^2 H^2 + \frac{1}{a^2} \right)
, \hspace{1cm} S_{23}=0.
\end{eqnarray*}
Thus, it is clear that the Einstein field equations (\ref{a}), (\ref{b})
and (\ref{c}) are satisfied. However this metric {\it does not} represent
a perfect fluid because the fluid velocity vector is not defined. The reason
is that the quadratic equation (\ref{a}) states exactly that the determinant
of the $2 \times 2$ matrix
\begin{center}
$ \left (
\begin{array}{cc}
S_{00}+S_{22} & S_{01} \\
S_{01} & S_{11}-S_{22} 
\end{array}
\right ) $
\end{center}
vanishes. This is in turn equivalent to the existence of two functions
$U_0$ and $U_1$ such that
\begin{center}
$\left (
\begin{array}{cc}
S_{00}+S_{22} & S_{01} \\
S_{01} & S_{11}-S_{22}
\end{array}
\right ) = \sigma \left (
\begin{array}{cc}
U_0^2 & U_0 U_1 \\
U_0 U_1 & U_1^2
\end{array}
\right ),$ 
\end{center}
where $\sigma$ is the sign of the component $S_{00}+S_{11}$ (if it
vanishes then $\sigma$ is the sign of $S_{11}-S_{22}$, if both
are zero then the whole matrix is vanishing). When these two functions
$U_0$ and $U_1$ satisfy
\begin{eqnarray*}
U_0^2- U_1^2 > 0
\end{eqnarray*}
we can normalise this ``vector" and define the two components $u_0$ and
$u_1$ of the 
fluid velocity vector as 
\begin{eqnarray*}
\sigma U_0^2 \equiv \left (\rho + p \right) u_0^2, 
\hspace{2cm}
\sigma U_1^2  \equiv \left (\rho + p \right) u_1^2, \hspace{1cm}
\mbox{with} \hspace{1cm} u_0^2 - u_1^2 = 1.
\end{eqnarray*}
These expressions also define $\rho + p$. 
However, other possibilities may also happen. First, there exist solutions
of the quadratic equation (\ref{a}) which satisfy
\begin{eqnarray*}
U_0^2 - U_1^2 < 0.
\end{eqnarray*}
This case would correspond to a perfect-fluid-like Einstein tensor but
with the ``fluid" velocity being spacelike and thus it has no physical interest
as it does not describe any known matter. Finally, there exist solutions
satisfying
\begin{eqnarray*}
U_0^2 - U_1^2 = 0 , \hspace{1cm} 
\Longleftrightarrow \hspace{1cm} S_{00}+S_{22} = S_{11}- S_{22} .
\end{eqnarray*}
In this case the Einstein tensor is of the form
\begin{eqnarray*}
S_{\alpha\beta} = K_{\alpha} K_{\beta} + p g_{\alpha \beta}
\end{eqnarray*}
where $\vec{K}$ is a null vector lying in the two plane spanned by the Killings
$\partial_t$ and $\partial_{\phi}$. Thus, this solution represents a radiative
fluid with pressure. So, we conclude that the
quadratic equation (\ref{a}) contains solutions which do not represent
perfect-fluids and that this must be checked after having found the solution.
It is clear now that the solution (\ref{solp12}) belongs to this class of
radiative fluids with pressure 
and thus will not be considered here any further. 
We can
now move to the following solution given by

\paragraph{Solutions 2 and 3}
Let us define the function
\begin{eqnarray*}
W(x,y) \equiv r(x) - \frac{y}{\alpha^2}
\end{eqnarray*}
so that $\Psi$ for Solution 2 is written as
$\Psi = A e^{\alpha r(x) + \frac{y}{\alpha}} W$.
The calculation of the vector (\ref{vec}) gives now
$\vec{S} = \Psi  A e^ {\alpha r(x) + \frac{y}{\alpha}} \left[ W \vec{Y_1} +  
\vec{Y_2} \right]$
where the two vectors $\vec{Y_1}(x)$ and $\vec{Y_2}(x)$ are given by
the expressions
\begin{eqnarray*}
\vec{Y_1} = \left (  R_1 + \alpha r' R_2 + \frac{2}{\alpha^2} + 
\frac{2c}{\alpha},  M_1 + \alpha r' M_2- \frac{2}{\alpha_1},
L_1 + \alpha r' L_2 + \frac{L_3}{\alpha} \right ), \\
\vec{Y_2} = \left ( r' R_2 - \frac{4}{\alpha^3} - \frac{2c}{\alpha^2},
r' M_2 + \frac{4}{\alpha^3},r' L_2  - \frac{L_3}{\alpha^2} \right ).
\hspace{15mm}
\end{eqnarray*}
From the fact that $W$ depends on $y$ while the two vectors
$\vec{Y_1}$ and $\vec{Y_2}$ involve only functions of $x$ it follows
that the vanishing of the norm of the vector $\vec{S}$ is equivalent to the
three equations:
\begin{equation}
 \vec{Y_1} \cdot \vec{Y_1}  = 0, \hspace{1cm}
 \vec{Y_1} \cdot \vec{Y_2}  = 0, \hspace{1cm}
 \vec{Y_2} \cdot \vec{Y_2} = 0. \label{sol2}
\end{equation}
This is a system of ordinary differential equations to be satisfied by the
unknown metric coefficients. At this stage, however, it is convenient to study
the possibility given by Solution 3 rather than studying this system of
equations because, as we will see, the two systems of equations for Solutions
2 and 3
can be treated together.

For Solution 3 it is worth defining the two
functions
\begin{eqnarray*}
\Psi_1(x,y) = e^{\alpha_1 r(x) + \frac{y}{\alpha_1}}, \hspace{1cm}
\Psi_2(x,y) = e^{\alpha_2 r(x) + \frac{y}{\alpha_2}},
\end{eqnarray*}
so that $\Psi$ is written as $\Psi = A \Psi_1 + B \Psi_2$,
where $A$ and $B$ are non-vanishing constants and $\alpha_1 \neq \alpha_2$.
Taking the partial derivatives of the function $\Psi$, the vector
(\ref{vec}) takes the form $\vec{S} = \Psi \left[ A \Psi_1 \vec{Z_1} +
B \Psi_2\vec{Z_2} \right]$,
where the two vectors $\vec{Z_1}(x)$ and $\vec{Z_2}(x)$ are
\begin{eqnarray*}
\vec{Z_1} = \left ( R_1 + \alpha_1 r' R_2 + \frac{2}{\alpha_1^2} + 
\frac{2c}{\alpha_1} , M_1 + \alpha_1 r' M_2- \frac{2}{\alpha_1^2} ,
L_1 + \alpha_1 r' L_2 + \frac{L_3}{\alpha_1}  \right ), \\
\vec{Z_2} = \left ( R_1 + \alpha_2 r' R_2 + \frac{2}{\alpha_2^2}
+ \frac{2c}{\alpha_2}, M_1 + \alpha_2 r' M_2- \frac{2}{\alpha_2^2},
L_1 + \alpha_2 r' L_2 + \frac{L_3}{\alpha_2} \right ). 
\end{eqnarray*}
As $\Psi_1^2$, $\Psi_1 \Psi_2$ and $\Psi_2^2$ are 
linearly independent functions of $y$, it follows that the quadratic equation (\ref{a})
is equivalent to the vanishing of all the scalar products of $\vec{Z_1}$ and
$\vec{Z_2}$ so that both vectors must be proportional to some null
vector. Of course any linear combination of them has also the same
property. It is convenient to consider the linear combination given by
\begin{eqnarray*}
\vec{Z_1}-\vec{Z_2}= \left[\alpha_1 - \alpha_2\right]
\left(r' R_2 - \frac{2 \alpha_1 +2 \alpha_2}
{\alpha_1^2 \alpha_2^2}-\frac{2c}{\alpha_1 \alpha_2},
r' M_2 + \frac{2\alpha_1 +2 \alpha_2}{\alpha_1^2\alpha_2^2},
r' L_2 - \frac{L_3}{\alpha_1 \alpha_2}\right).
\end{eqnarray*}
As $\alpha_1 - \alpha_2$ is non-vanishing, we can drop this global factor
and the full set of Einstein's field equations to be satisfied in this case are
given by 
\begin{eqnarray}
r'' + \left (\alpha_1 + \alpha_2 \right ) {r'}^2 + 
\frac{\alpha_1+ \alpha_2}{\alpha_1^2 \alpha_2^2} =0, \label{Eq0} \hspace{4cm}\\
H=r', \label{Eq23} \hspace{7cm}\\
G + \alpha_1 \alpha_2 {r'}^2 + \frac{ \left ( \alpha_1+ \alpha_2
\right )^2}{\alpha_1^2 \alpha_2^2} - \frac{1}{\alpha_1\alpha_2} = 0,  
\label{Eq2233} \hspace{30mm} \\
\left ( R_1 + \alpha_1 r' R_2 + \frac{2}{\alpha_1^2} + \frac{2c}{\alpha_1}
\right ) \left ( M_1 + \alpha_1 r' M_2 - \frac{2}{\alpha_1^2} \right ) - 
\left ( L_1 + \alpha_1 r' L_2 + \frac{L_3}{\alpha_1} \right )^2 = 0, \label{Eqa} \\
\left (R_1 + \alpha_1 r' R_2 + \frac{2}{\alpha_1^2} + \frac{2c}{\alpha_1} 
\right ) \left ( r' M_2 + 2 \frac{\alpha_1+ \alpha_2}{\alpha_1^2 \alpha_2^2} 
\right )
+ \left ( r' R_2 - 2 \frac{\alpha_1+ \alpha_2}{\alpha_1^2 \alpha_2^2} -\right.
\hspace{1cm} \nonumber \\
\left . - \frac{2c}{\alpha_1\alpha_2} \right )
 \left ( M_1 + \alpha_1 r' M_2 - \frac{2}{\alpha_1^2} \right ) 
-  2 \left ( L_1 + \alpha_1 r' L_2 +
\frac{L_3}{\alpha_1 } \right )
\left ( r' L_2 - \frac{L_3}{\alpha_1\alpha_2} \right )=0 ,  \label{Eqb}\\
\left ( r' R_2 - 2 \frac{\alpha_1+ \alpha_2}{\alpha_1^2\alpha_2^2} - \frac{2c}{
\alpha_1\alpha_2} \right )
\left ( r' M_2 + 2 \frac{\alpha_1 + \alpha_2}{\alpha_1^2\alpha_2^2}\right )
- \left ( r' L_2 - \frac{L_3}{\alpha_1\alpha_2} \right )^2=0 . \label{Eqc}
 \hspace{7mm}
\end{eqnarray}
At the sight of this system of ordinary differential equations, we note that
although Solution 3 is characterized by the property
$\alpha_1\neq \alpha_2$, the system (\ref{Eq0})-(\ref{Eqc}) itself can
 be particularized to
$\alpha_1=\alpha_2$ without problem (this is due to the fact that
we have dropped a global factor $\alpha_1-\alpha_2$ above) and gives
exactly the system (\ref{sol2}) which had to be satisfied for the previous
Solution 2. Thus, by studying the compatibility of system
(\ref{Eq0})-(\ref{Eqc}) we are
considering both Solutions 2 and 3 together and thereby we have reduced
considerably the amount of work to be done. By assuming that the resulting
spacetime has a well-defined axis of symmetry we can expand the functions
involved in this system around the axis of symmetry and check the
compatibility of this system of differential equations. In the
Appendix we show that  
that the system (\ref{Eq0})-(\ref{Eqc}) has no solution giving rise to
a spacetime with regular axis of symmetry.
Let us emphasize that this assumption of regular axis is not an additional
condition in our work because the restriction of the possible Lie algebras
to the four cases listed in the Introduction is based on a geometrical
result proven in \cite{S2} which makes essential use of this regularity
property.

\paragraph{Solution 4}
This last case is the most interesting one because it gives rise to a family
of stationary non-convective differentially rotating perfect-fluid solutions.
In fact, from the explicit form of $\Psi = A e^{\alpha r(x) + \frac{y}{\alpha}}$
the vector (\ref{vec}) takes the form
\begin{eqnarray*}
\vec{S} = A \Psi^2 \left ( R_1 + \alpha r' R_2 + \frac{2}{\alpha^2} + 
\frac{2c}{\alpha} , M_1 + \alpha r' M_2- \frac{2}{\alpha^2} ,
L_1 + \alpha r' L_2 + \frac{L_3}{\alpha}  \right ),
\end{eqnarray*}
so that the condition that this vector is null gives only one condition
for the three functions $F$, $Q$ and $P$. Besides, we know that for this
Solution 4, $H$ and $G$ had to satisfy only the two relations $H=r'$ and
(\ref{apCG}), with no relation involving only $r$. Thus, the Einstein
field equations reduce to a system of {\it three} ordinary
differential equations for the {\it four} unknowns $r$, $F$, $Q$ and $P$.
We devote the
following section to consider these perfect-fluid solutions.

\section{The solutions}
In order to write down explicitly the field equations and the density
and pressure for the family just found, let us rewrite the line-element in a
separate way for the Cases II and III. It is obvious that the metric for the
Case II can be rewritten as
\begin{eqnarray}
ds^2= e^{2bcy}\left[ -F(x)\left( \frac{}{} e^{-c y}
dt+ P(x) d\phi \right )^2+ \frac{Q^2(x)}{F(x)}d\phi^2+
\frac{1}{L^2(x)}dx^2+\frac{1}{L^2(x)}dy^2 \right ] \label{leII}
\end{eqnarray}
where $b$ is a constant and the functions $F$, $Q$ and $P$ are {\it
different}
from those used in the previous sections. 
We
do not use different symbols for this metric coefficients in order to
avoid using too much notation.
In terms of these functions, the field equations to be
solved read
\begin{eqnarray}
\f+ \frac{F^2}{Q^2} PP' + \left (2-4b\right) \ele =0,  \label{ss23}
 \hspace{45mm}\\
\frac{1}{2} \Qf - \f\q + 2 \ele \q - \frac{1}{2} \frac{F^2}{Q^2} {P'}^2 + \Dq
+ \frac{c^2}{2} \frac{P^2F^2}{Q^2} +c^2\left(2b^2-1\right) = 0, \label{ss2233} 
\hspace{1cm} \\
\left [ \frac{1}{2} \left ( \f \right )'+ \frac{1}{2} \f \q + \left ( \ele 
\right )' +  \ele \q + \frac{1}{2} \frac{F^2}{Q^2} {P'}^2 +2bc^2\left(b-1\right)
+ c^2 \frac{P^2F^2}{Q^2} \right ] \hspace{1cm} \nonumber\\
\left [ \frac{1}{2} \left ( \f \right )'
+ \frac{1}{2} \f \q - \left ( \ele\right )' - \ele \q+ 
\frac{1}{2} \frac{F^2}{Q^2} {P'}^2 - \Dq -c^2\left(2b^2-1\right) \right ] = 
\hspace{8mm} \nonumber \\
= \frac{F^2}{Q^2} \left [\f P' + \frac{1}{2} P'' - \frac{1}{2}
\q P' + c^2P\left(b-1\right) \right ]^2. \label{sseq}
\end{eqnarray}
The explicit expressions for the density and pressure are simply
\begin{eqnarray*}
p= \frac{L^2}{e^{2bcy}} \left [ - \frac{1}{4} \Qf + \frac{1}{2} \f\q
- \ele \q + \frac{1}{4} \frac{F^2}{Q^2} {P'}^2 -
\frac{c^2}{4} \frac{P^2 F^2}{Q^2} + c^2\left(b-1\right)^2 \right ], \nonumber \\
\rho = \frac{L^2}{e^{2bcy}} \left [ 2 \left ( \ele \right )'
+ 2 \ele \q + \frac{1}{2} \Dq + c^2\frac{P^2 F^2}{Q^2}
+c^2\left(2b^2- \frac{3}{2}\right) \right ], \hspace{1cm}
\end{eqnarray*}
while the fluid velocity vector of this solution reads
\begin{eqnarray*}
\vec{u} = e^{-bcy}A(x) \left ( e^{cy} \frac{\partial}{\partial t} +
R(x) \frac{\partial}{\partial \phi} \right ),
\end{eqnarray*}
where $A(x)$ and $R(x)$ are rather long expressions depending on the
functions $F$, $Q$, $P$ and $L$. We note that the solution is
differentially rotating except when $R(x) \equiv 0$, which corresponds
to the case when both the second and third terms in
square brackets in the last equation are vanishing. 

The system of equations (\ref{ss23})-(\ref{sseq}) is formed
by three ordinary differential equations for four unknowns so that, in
principle, it contains an arbitrary function in its general solution. However,
the third equation in the system is quadratic so that the compatibility
of the system is not obvious at first sight. In order to see explicitly
that the system is indeed compatible and contains an arbitrary function,
let us define a new function $M(x)$ by
\begin{eqnarray*}
F \equiv \frac{Q \sqrt{M}}{P}.
\end{eqnarray*}
Thus, we are assuming $P\neq 0$ (the case $P\equiv 0$ will be
dealt with below). Notice that when $P$ is non-zero we can always make it
positive (locally) by using the coordinate change $\phi \rightarrow - \phi$.
Now, we define three new functions $u(x)$, $s(x)$ and $w(x)$ by
\begin{eqnarray*}
\frac{Q'}{Q} \equiv u, \hspace{1cm} \frac{P'}{P} \equiv s, \hspace{1cm}
\frac{L'}{L} \equiv w.
\end{eqnarray*}
We will consider $M$ as the arbitrary function so that the system gives
differential equations for $u$, $s$ and $w$. First of all we notice that the
case  $M \equiv 1$ gives a solution which is not a perfect fluid (see the
discussion above) and, therefore, we can assume $M \neq 1$ from now on.
From equation (\ref{ss23}) we obtain $u$ and substitute it into the other two
equations. Then, (\ref{ss2233}) allows us to obtain $s'$ in terms of the
rest of the functions. Substituting into (\ref{sseq})
we obtain an expression of the form
\begin{eqnarray*}
{w'}^2 \left ( M-1 \right ) \left (4b^2- 4b + M \right)
+ w' \Sigma_1 \left (M,M'M'', w,s \right) + \Sigma_2 \left (M,M',M'',w,s
\right) =0.
\end{eqnarray*}
When $M \equiv 4b - 4 b^2$ the system (\ref{ss2233})-(\ref{sseq})
is linear in their two highest derivatives
$(w',s')$ and therefore the existence theorems apply. In the more general
case $M\neq 4b - 4b^2$, we have to solve $w'$ from this quadratic equation
to obtain a system which is linear in their highest derivatives. This can be
done unless the discriminant of this equation is negative. Evaluating this
discriminant we notice that contains both positive and negative terms so
that it is not negative definite for all $M$ and all initial conditions. 
In general, there will be  functions $M$ and initial conditions for
$n$ and $s$ (as well as for the constant $b$)
which make the discriminant positive thus giving solutions to
the system. Just to give an explicit very particular example
(they are trivial to construct), the value $b=1$,
the initial conditions $w\mid_{x=0} =0$,
$s \mid_{x=0}$ and any function
M with $M \mid_{x=0} =1/2,$ $M' \mid_{x=0}=0$
and $M'' \mid_{x=0} =0 $ give a positive discriminant and produce a
well-defined solution. Thus, we conclude that, even though we cannot prescribe
the function $M$ freely, the general solution of the system
(\ref{ss23})-(\ref{sseq}) contains an arbitrary function $M$ belonging to a
suitable open subset of the whole class of $C^2$ functions. 

Let us now consider what happens when $P \equiv 0$ (which corresponds
to the static limit of the family).
It is indeed very simple to show
that the equations can be rewritten in this case in the form
\begin{eqnarray}
\frac{L'}{L} \equiv w(x) \neq 0
, \hspace{1cm} \frac{F'}{F} = 2 \left (2b-1 \right )
w, \hspace{1cm} \frac{Q'}{Q} = \frac{w'}{w} + \frac{\left (2b-1\right)^2}{b-1}
w, \hspace{1cm} (b \neq 1) \nonumber \\
\frac{w''}{w} + \frac{8b^2 - 4b -1}{b-1} w' + \frac{\left (2 b^2 -1
\right )^2}{
\left (b-1 \right)^2} \left [ \left (2b-1 \right)^2 w^2 + \left (b-1 \right)^2
 c^2 \right ] = 0 \hspace{1cm} \label{pezero}
\end{eqnarray}
(the case $w=0$ is slightly different and can be integrated completely to give
a solution which contains more symmetry). The energy-momentum contents
for the solutions of (\ref{pezero}) is that of a fluid with velocity vector
proportional to $\partial_t$ and density and pressure given by
\begin{eqnarray*}
\rho = \frac{L^2}{e^{2bcy}} \frac{\left( b+1 \right) }{\left( b-1\right) }
\left [ 2 \left (b-1 \right) w'
+ \left (2b-1 \right)^2 w^2 + \left (b-1 \right)^2 c^2 \right ], \hspace{1cm}
p = \frac{b-1}{b+1} \rho,
\end{eqnarray*}
so that it satisfies a barotropic equation of state.

We consider now the  Case III. The metric takes the form
\begin{eqnarray}
ds^2= e^{2bvy}\left[ -F(x) dt^2
+ \frac{Q^2(x)}{F(x)} \left (\frac{}{}d\phi + \left(P(x)-v y\right ) dt \right 
)^2 +\frac{1}{L^2(x)}\left(dx^2+dy^2\right) \right ] \label{leIII}
\end{eqnarray}
where the functions $F$ and $Q$ have been again 
redefined by introducing $L(x)$.
Einstein's equations with these new functions are 
\begin{eqnarray}
b \ele + \frac{1}{4} \frac{Q^2}{F^2} P' = 0, \hspace{55mm} \label{sss23}\\
\frac{1}{2} \Qf - \f \q + 2 \ele \q - \frac{1}{2}\frac{Q^2}{F^2} {P'}^2
+ \Dq + 2 b^2 v^2 + \frac{v^2}{2} \frac{Q^2}{F^2} = 0, \hspace{17mm}
\label{sss2233}  \\
\left [ \frac{1}{2} \left ( \f \right )' + \frac{1}{2} \f \q +
\left ( \ele \right )' + \ele \q - \frac{1}{2} \frac{Q^2}{F^2} {P'}^2 + 2b^2v^2
\right ] \left [\frac{1}{2} \Df - \frac{1}{2} \f\q - \right . \nonumber \\
\left . -\left ( \ele \right )' +\ele \q - \frac{Q^2}{F^2} {P'}^2 -
\frac{v^2}{2}\frac{Q^2}{F^2} \right ] = \frac{Q^2}{F^2} \left [ \f P'
- \frac{1}{2} P'' - \frac{3}{2} \q P' + b v^2 \right ] ^2. \label{ssseq} \\
\end{eqnarray}
The fluid velocity vector of this solution reads
\begin{eqnarray*}
\vec{u} = e^{-bvy} A(x) \left ( \frac{\partial}{\partial t} + \left [
vy + R(x) \right ] \frac{\partial}{\partial \phi} \right )
\end{eqnarray*}
where, as before, $A(x)$ and $R(x)$ are expressions depending on
the four functions $F$, $Q$, $P$ and $L$. We note that in this
case the solution is {\it always} differentially rotating (its
rigidly rotating limit must satisfy $v=0$ and thus it actually belongs to the
case when the conformal Lie algebra is abelian).
The expressions for the density and pressure for this solution read
\begin{eqnarray*}
p = \frac{L^2}{e^{2bvy}} \left [  - \frac{1}{4} \Qf + \frac{1}{2} \f \q
- \ele \q + \frac{1}{4} \frac{Q^2}{F^2} {P'}^2 + b^2 v^2 - \frac{v^2}{4}
\frac{Q^2}{F^2} \right ], \\
\rho = \frac{L^2}{e^{2bvy}} \left [ 2 \left (\ele \right )' + 2 \ele \q
+ \frac{1}{2} \Dq + 2 b^2 v^2 + v^2 \frac{Q^2}{F^2} \right ]. \hspace{1cm}
\end{eqnarray*}

The same considerations we made regarding the existence of solutions for the
qua\-dra\-tic set of differential equations (\ref{ss23})-(\ref{sseq}) hold
in this case. Given that the equations are  simpler, the analysis
is easier than before. We simply need to define a new function $N(x)$ by
\begin{eqnarray*}
Q \equiv F N
\end{eqnarray*}
and regard $N$ as the arbitrary function in the system. Introducing two
new functions $s(x)$ and $w(x)$ by
\begin{eqnarray*}
\frac{F'}{F} \equiv s, \hspace{1cm} \frac{L'}{L} \equiv w,
\end{eqnarray*}
we notice that (\ref{sss23}) allows us to obtain $P'$ in terms of $N$ and $w$.
Substituting into the other two equations we find that (\ref{sss2233}) contains
$s'$ linearly and no other derivatives (except for $N$).
Substituting $s'$ obtained from this equation into the last one we
find, similarly as as before,
\begin{eqnarray*}
\left ( 1 + \frac{4b^2}{N^2} \right) {w'}^2 + \Sigma_1 \left (N,N',N'',w,s 
\right)
w' + \Sigma_2 \left (N,N',N'',w,s \right) = 0.
\end{eqnarray*}
Evaluating the discriminant of this quadratic equation it can be seen that
it is not negative
definite. Thus, there exist initial conditions and functions $N$ which make it
positive. A trivial example is given now by
$s \mid_{x=0} = 0$, $w \mid_{x=0} =0$, $b= 1/2$ and any function $N$ satisfying
$N \mid_{x=0} =1$, $N' \mid_{x=0} =0$ and $N'' \mid_{x=0} =0$. Therefore,
the general solution of the system (\ref{sss23})-(\ref{ssseq}) will contain
an arbitrary function belonging to some open non-empty subset of the
class of $C^2$ functions.

\vspace{3mm}

It is interesting to note also that the conformal Killing of
all these solutions (both for algebra II and algebra III)
is in fact homothetic as it satisfies
\begin{eqnarray*}
{\cal{L}}_{\vec{k}}\,g_{\mu\nu} = 2 bc g_{\mu\nu} \hspace{1cm} \mbox{(Case II)},
\hspace{15mm}
{\cal{L}}_{\vec{k}}\,g_{\mu\nu} = 2 bv g_{\mu\nu} \hspace{1cm} \mbox{(Case III)},
\end{eqnarray*}
so that the scale factor is a constant. Thus, the barotropic equation of state
of the perfect fluid, in case it exists, must be a linear relation of the form 
$p = \left (\gamma +1 \right ) \rho$ for some constant $\gamma$ (see \cite{Wn}).
As the families of solutions contain an arbitrary function, there always exist
solutions satisfying this barotropic equation of state. 

Let us finally
summarize the main results of this paper and Ref.\cite{S1} in the
following table where the solutions of Einstein's field equations for
stationary and axisymmetric space-times with a (non-isometric) conformal motion
(proper for the Abelian Case and Case I, but homothetic for the remaining Cases
II and III) and filled with a non-convective perfect fluid are written down. 

\begin{center}
\begin{tabular}{|c||l@{}|}
\hline
Abelian Case & \begin{tabular}{@{}l|l@{}}
                ${\Psi}_{,x} \neq 0 $ \hspace{14.5mm} & Schwarzschild interior
                solution \\
                & \\
                ${\Psi}_{,x} = 0$ & Solution in \cite{S3} 
                \end{tabular}
\\
\hline
Case I & \begin{tabular}{@{}l|l@{}}
         ${\Psi}_{,x} \neq 0 $ \hspace{14.5mm} & Schwarzschild interior
                                                solution\\ 
          & \\
         ${\Psi}_{,x} = 0$ & Solution in \cite{S1}  
          \end{tabular}
\\
\hline
Case II & \begin{tabular}{@{}l|l@{}}
           & \\
          ${\Psi}_{,y} = -bc \Psi$ \hspace{6.7mm} & Solution (\ref{leII})
           \hspace{3mm}\\
           & 
          \end{tabular}
\\
\hline
Case III & \begin{tabular}{@{}l|l@{}}
            & \\
           ${\Psi}_{,y} = -bv \Psi$ \hspace{6.3mm} & Solution (\ref{leIII})
           \hspace{3mm} \\
            & 
            \end{tabular}                 
\\
\hline
\end{tabular}
\end{center}

\subsection*{Corrigendum to paper \cite{S1}}
Let us end this paper by adding a corrigendum to our previous paper \cite{S1}.
The general solution for the Case I was explicitly found there. However, the
expression for the fluid velocity vector (the expression following formula (27)
in \cite{S1}) is wrong and should read
\begin{eqnarray*}
\vec{u} = \frac{1}{\beta \sqrt{ \dot{H}^2 - H^2 \omega^2 X^2}}
\frac{\partial}{\partial t }
\end{eqnarray*}
which implies that the rotation $\Omega$ of the fluid is constant.
Thus the general perfect fluid solution in Case I is {\it rigidly rotating}
contrarily to what was stated in \cite{S1}. As a consequence, we have that
this solution must be contained in the general Wahlquist family \cite{Wa} which
is the general solution (together with its limit cases
\cite{Kr1}\cite{Kr2}\cite{Sen})
for Petrov type D stationary and axisymmetric rigidly rotating
perfect fluids with equation of state $\rho+3p=$const. (see \cite{Sen}). 

\section*{Acknowledgements}
The present work has been partially supported by the Spanish
Ministerio de Educaci\'on y Ciencia under project No. PB93-1050.
M. Mars wishes to thank the Ministerio de Educaci\'on y Ciencia
for financial support.

All the tensors in this paper have been computed with the
algebraic computer programs CLASSI and REDUCE. The calculations 
involved in the proof of incompatibility of the system of
differential equations (\ref{Eq0})-(\ref{Eqc}) have
been performed with the essential help of REDUCE.

\section{Appendix}
The aim of this Appendix is to show that the overdetermined system
of ordinary differential equations (\ref{Eq0})-(\ref{Eqc})
has no solution representing an
axially symmetric spacetime with a well-defined axis of symmetry. 
First of all, we need to find the conditions on the metric functions $F$,
$P$ and $Q$ such that the metric (\ref{metr}) has well-defined
axial symmetry. Thus, we must impose the existence of a two-dimensional
surface where the axial Killing vector $\partial_\phi$ vanishes
and where the so-called elementary flatness condition (see e.g.
\cite{KMSH}) is fulfilled. A trivial calculation shows that these
conditions are equivalent to the existence of a finite value of $x$
(which can be redefined to $x=0$ using a coordinate change $x \longrightarrow
x + \mbox{const.}$) where 
\begin{eqnarray}
Q \mid_{x=0} = 0 , \hspace{1cm} F \mid_{x=0}=1 \hspace{1cm} \mbox{and}
\hspace{1cm} Q' \mid_{x=0} = 1 \label{condition}
\end{eqnarray}
(in order to find these expressions we have also made use of the
coordinate freedom of rescaling the coordinate $t$).
We must now
expand the metric functions around $x=0$ and show that the system
of ordinary equations (\ref{Eq0})-(\ref{Eqc}) is incompatible. The first
thing we notice is that the equation (\ref{Eq0}) for $r$ can be
completely integrated to give
\begin{eqnarray*}
r' = -\frac{1}{\alpha_1 \alpha_2} \tan \left [ \frac{\alpha_1 + \alpha_2}{
\alpha_1 \alpha_2} \left (x-x_0 \right) \right ]
\hspace{1cm} \mbox{if} \hspace{5mm} \alpha_1 + \alpha_2 \neq 0, \\
r' = \mbox{const} \neq 0 \hspace{25mm} \mbox{if} \hspace{5mm}
\alpha_1+\alpha_2 =0.
\end{eqnarray*}
In the second case we assume the constant $r'$ to be non-zero because
otherwise the equation (\ref{Eq23}) gives $H=0$ which belongs to another
case. Now,
a very simple analysis of the function $\Psi$ (which is given
by Solution 2 or Solution 3) shows that $r'$ 
must be finite at $x=0$ in order to obtain
a non-singular metric at the axis of symmetry $x=0$. This implies that
$\frac{\alpha_1+\alpha_2}{\alpha_1\alpha_2} x_0 \neq \pm \left ( \frac{\pi}{2}
+ \pi n \right )$ , ($n$ being an integer). This result is necessary
in order to consider properly the different powers in $x$ of the equations.

We must now consider
separately  
the case III from the case II.
The first one $(c=0, v \neq 0)$ is simpler to study and the proof
is as follows. First, we
obtain 
\begin{eqnarray*}
P' = \frac{4 r'}{v } \frac{F^2}{Q^2}
\end{eqnarray*}
from the equation (\ref{Eq23}) and substitute this expression everywhere
so that only the functions $F$, $Q$ and $r'$ are present in the equations.
We define a new function $w(x)$ by $Q \equiv x w(x)$ so that
$w(0)=1$ (see (\ref{condition})). The coefficient in $x^{-2}$ of equation (\ref{Eqc}) gives
simply
\begin{eqnarray*}
-\frac{16}{v^2} {r'}^4 \mid_{x=0}= 0,
\end{eqnarray*}
which implies from the expression for $r'$ above that 
$x_0 = 0$ and $\alpha_1 + \alpha_2 \neq 0$. Now, 
the coefficients in $x^{-1}$ of equation (\ref{Eq2233}) and (\ref{Eqb})
read, respectively
\begin{eqnarray*}
-\frac{1}{2} \left [ F' \mid_{x=0} - 2 w' \mid_{x=0} \right ] = 0, 
\hspace{1cm}
- \frac{\alpha_1+\alpha_2}{\alpha_1^2 \alpha_2^2} \left [ F' \mid_{x=0}
-4 w' \mid_{x=0} \right ] =0 
\end{eqnarray*}
so that obviously $F'\mid_{x=0}=0$ and $w' \mid_{x=0}=0$. Using these
expressions
the coefficient in $x^{-2}$ of (\ref{Eqa}) reads
\begin{eqnarray*}
-\frac{16}{v^2}
\frac{\left( \alpha_1 + \alpha_2 \right)^2}{\alpha_1^4 \alpha_2^4} = 0
\end{eqnarray*}
which is impossible and we have shown that no regular solutions around
the axis of symmetry exist for the case III.  

Regarding the case II $(c \neq 0, v=0)$, the proof is more involved
because the equations are more complicated. The idea is, however,
the same as in the previous case. Now it is convenient to
define a new function $M(x)$ by 
\begin{eqnarray}
F = \frac{PQ}{\sqrt{M}}, \label{eem}
\end{eqnarray}
which allows us to write (\ref{Eq23}) in the form
\begin{eqnarray}
\frac{P'}{P} = \frac{1}{M-1} \left ( -\frac{1}{2} \frac{M'}{M} +
\frac{Q'}{Q} + \frac{4 r'}{c} \right ) \label{subst}
\end{eqnarray}
(the subcase $M \equiv 1$ can be easily shown to be impossible) and substitute
$\frac{P'}{P}$ everywhere so that only the functions $Q$, $M$ and $r'$ appear
in the system. We have to take into account, however, that we still do not
know the behaviour of the function $M$ near $x=0$ because the regularity
on the axis of symmetry does not tell us anything about the function $P$
(and therefore about $M$ through its definition), except that $QP$ must be
finite at $x=0$ (because the norm of the timelike Killing $\partial_t$
must be finite and negative on the axis of symmetry) and, therefore,
the function $M$ must be regular at $x=0$. Our next aim is to show that
necessarily $r' \mid_{x=0} =0$ (or, equivalently, $x_0 =0$ and 
$\alpha_1+\alpha_2 \neq 0$). Let us suppose, on the contrary, that 
$r' \mid_{x=0} \neq 0$ (but finite as discussed above). Again we define the
function $w(x)$ by $Q \equiv x w(x)$ and substitute everywhere.
The coefficient in $x^{-2}$
of (\ref{Eqc}) gives
\begin{eqnarray*}
- M \mid_{x=0} {r'}^2 \mid_{x=0} = 0 \hspace{1cm} \Longrightarrow 
\hspace{1cm} M \mid_{x=0} =0
\end{eqnarray*}
and using this information, the coefficient in $x^{-3}$ of (\ref{Eqa})
gives $M' \mid_{x=0} = 0 $ and then the coefficient in $x^{-2}$ in
(\ref{Eqb}) gives
$M'' \mid_{x=0}=0$. Thus, we have that $M$ has a zero of
order $n \geq 3$ at $x=0$ which implies from its definition (\ref{eem}) that
$P$ has a zero of order $\frac{n}{2} -1$, but then the expression 
(\ref{subst}) is impossible in the limit $x=0$ and we have proven that
$r' \mid_{x=0} =0$.

It is convenient to perform the change of variables
\begin{eqnarray*}
u =  - \tan \left [ \frac{\left (
\alpha_1 +\alpha_2\right) }{\alpha_1 \alpha_2} x \right ] ,
\end{eqnarray*}
so thar $r' = \frac{1}{\alpha_1 \alpha_2} u$ and the axis of symmetry
is located at $u=0$. Now, expanding $M$ and $w$ around $u=0$ it is not
difficult to show that the system of equations is incompatible. The
calculations are, however, long and the use of computer algebra
is needed. Moreover, a number of different subcases arise so that
the proof is lengthy and  cannot be reproduced here. Let us simply give
the main indications for the proof. It is enough to restrict the analysis
to the three simplest equations (\ref{Eq2233}), (\ref{Eqb}) and (\ref{Eqc})
and it is convenient to redefine the constants to
\begin{eqnarray*}
\alpha_1 \equiv \frac{1}{K-K_2}, \hspace{1cm} \alpha_2 \equiv \frac{1}{K_2}
\end{eqnarray*}
and define a function $n(x)$ through
\begin{eqnarray*}
\frac{Q'}{Q} = - \frac{\left (\alpha_1 + \alpha_2\right) }{\alpha_1^2 
\alpha_2^2} \frac{n}{r'}.
\end{eqnarray*}
Finally, the system of equations takes a much simpler form by performing
the following combination
\begin{eqnarray*}
& & \mbox{Equation1} \equiv 4 r'^2 \mbox{(\ref{Eq2233})} - 
 \mbox{(\ref{Eqc})},  \\
& & \mbox{Equation2} \equiv 2 {r'}^2 \mbox{(\ref{Eqb})} + r' \frac{d 
\mbox{(\ref{Eqc})}}{dx} -
2 \left [ {r'}^2 \left (\alpha_1 - \alpha_2 \right) + 
\frac{\alpha_1 + \alpha_2}{\alpha_1^2 \alpha_2^2} \left (n-1
\right) \right ] \mbox{(\ref{Eqc})},  \\
& & \mbox{Equation3} \equiv \mbox{(\ref{Eqc})}. 
\end{eqnarray*}
The first of these equations contains $M$ linearly and without derivatives
of $M$. This allows us to substitute $M$ from that equation into the two
others so that we have two equations for the single function  $n(u)$. The
analysis now is lengthy but trivial.

\end{document}